 \documentclass{appolb}
\usepackage{epsfig}
\usepackage{graphicx}
\newcommand{\beq}[2]{\begin{equation}#1\label{#2}\end{equation}}
\newcommand{\ceq}[1]{(\ref{#1})}

\begin{document}
\title{Dynamics of a three-dimensional
inextensible chain
\thanks{Talk presented by J. Paturej at the 21st Marian Smoluchowski Symposium on Statistical
Physics, Zakopane, Poland, September 13-18, 2008.
}%
}
\author{Franco Ferrari, Jaros{\l}aw Paturej
\address{Institute of Physics
and CASA*, University of Szczecin,
 Wielkopolska 15, 70-451 Szczecin, Poland}
\and Thomas A. Vilgis
\address{Max Planck Institute
for Polymer Research, 10
  Ackermannweg, 55128 Mainz, Germany}
}
\maketitle
\begin{abstract}
In  the first part of this work the
classical and statistical aspects of the
dynamics of
an inextensible chain
in three dimensions are investigated.
In the second part
the special case of a chain admitting only
fixed angles
with respect to the $z-$axis
is studied using a path integral approach.
It is shown that it is possible to
 reduce this problem to a two-dimensional case, in a way which is
 similar to the reduction of the
statistical mechanics of a directed polymer to the random walk of a
two-dimensional particle.
\end{abstract}

\section{Introduction}\label{introduction}

In this report we investigate the dynamics  of an
inextensible three-dimensional chain
fluctuating in some medium at fixed temperature $T$. The chain is
considered as the continuous
limit of a freely jointed chain, which consists of a set of $N-1$
rigid links of
length $a$ and $N$ beads of mass $m$
attached at the joints between two consecutive segments. The
formulation of the dynamics
of a chain with rigid constraints based on
 the stochastic equation of Langevin has been
extensively studied in a series of seminal papers by Edwards and Goodyear
\cite{EdwGoo1,EdwGoo2,EdwGoo3}. Unfortunately, to deal with these
constraints at the level of
stochastic equations
 is a cumbersome task.
Up to recent times, most of the developments
in the dynamics of a chain with rigid constraints have been confined
to numerical simulations,
see for example Refs.~\cite{ottinger,muthukumar,adland}.
For this reason, we recently proposed an interdisciplinary treatment
to this problem
combining methods of field
theory and statistical mechanics \cite{FEPAVI1}.
The strategy is to regard the change of the chain conformation
 as the
motion of Brownian particles with constrained trajectories. The
framework of the calculations is that of path integrals. The
constraints are introduced by a procedure which is commonly applied
in statistical mechanics in order to enforce topological conditions
on a system of linked polymers. One ends up in this way with a field
theory which is a generalized non-linear sigma model (GNL$\sigma$M).
Recently, this path integral formulation has been connected to the
usual description of the dynamics of a chain as a diffusion process
\cite{FPlangevin}. The GNL$\sigma$M may be applied to the cases of
an isolated cold chain or of a hot polymer in the vapor phase.
Applications of the GNL$\sigma$M have been developed in
Refs.~\cite{FEPAVI2,FEPAVI3}.
\\
This work is organized as follows. In Sec.~\ref{s:class} the dynamics of
a classical chain is investigated
in three dimensions. The kinetic energy of a discrete
chain with $N-1$
segments is derived
in cartesian and spherical coordinates. Moreover, the limit to a
continuous chain
is performed.
In Sec.~\ref{s:heatbath} the probability distribution function for an
inextensible chain in a heat
bath is constructed using a path integral approach.
Sec.~\ref{s:3dcb} is dedicated to the discussion of the dynamics of a
rigid chain
in which the segments are allowed to form only fixed angles
with respect to the $z$ axis.
Finally our Conclusions are drawn in Sec.~\ceq{conclusions}.

\section{Classical dynamics of a three-dimensional chain with rigid
  constraints}\label{s:class}

Let us consider a chain of $N-1$ segments $P_iP_{i-1}$ of fixed
lengths $l_i$ $(i=2,\ldots,N)$ embedded in a three-dimensional space.
With the symbol $l_1$ we denote the distance of the end point $P_1$
from the origin of the coordinate system.
Additionally,  there are small beads of mass $m_i$ attached at the joints of the segments $P_iP_{i-1}$,
where $i=1,\ldots,N$.

The above construction describes a {\it freely jointed random chain}, which
is one of the
basic models used in polymer physics. Freely jointed
means that a
given segment can take with equal probability any spatial
orientation independently of the orientations of the neighbouring
segments.
The position of each segment $P_iP_{i-1}$ can be specified by giving
the coordinates of its endings $P_i$ and $P_{i-1}$ in cartesian
coordinates
$P_i(t)=[x_i(t),y_i(t),z_i(t)]$.
However, in the following it will be more convenient to use spherical
coordinates:
\begin{eqnarray}
x_i(t)&=&\sum_{j=1}^il_j\cos\varphi_j(t)\sin\theta_j(t)\qquad\qquad
(i=1,\ldots,N)\nonumber\\
y_i(t)&=&\sum_{j=1}^il_j\sin\varphi_j(t)\sin\theta_j(t)\qquad\qquad
(i=1,\ldots,N)
\nonumber\\
z_i(t)&=&\sum_{j=1}^il_j\cos\theta_j(t)\qquad\qquad\qquad\qquad
(i=1,\ldots,N) \label{posns3d}
\end{eqnarray}
We will also neglect analytical complications connected
 with the inclusion of interactions
such as the hydrodynamic interaction and steric effects. In this sense
the chain is treated as a free one.

\begin{figure}[bpht]
\centering
\includegraphics[scale=0.25]{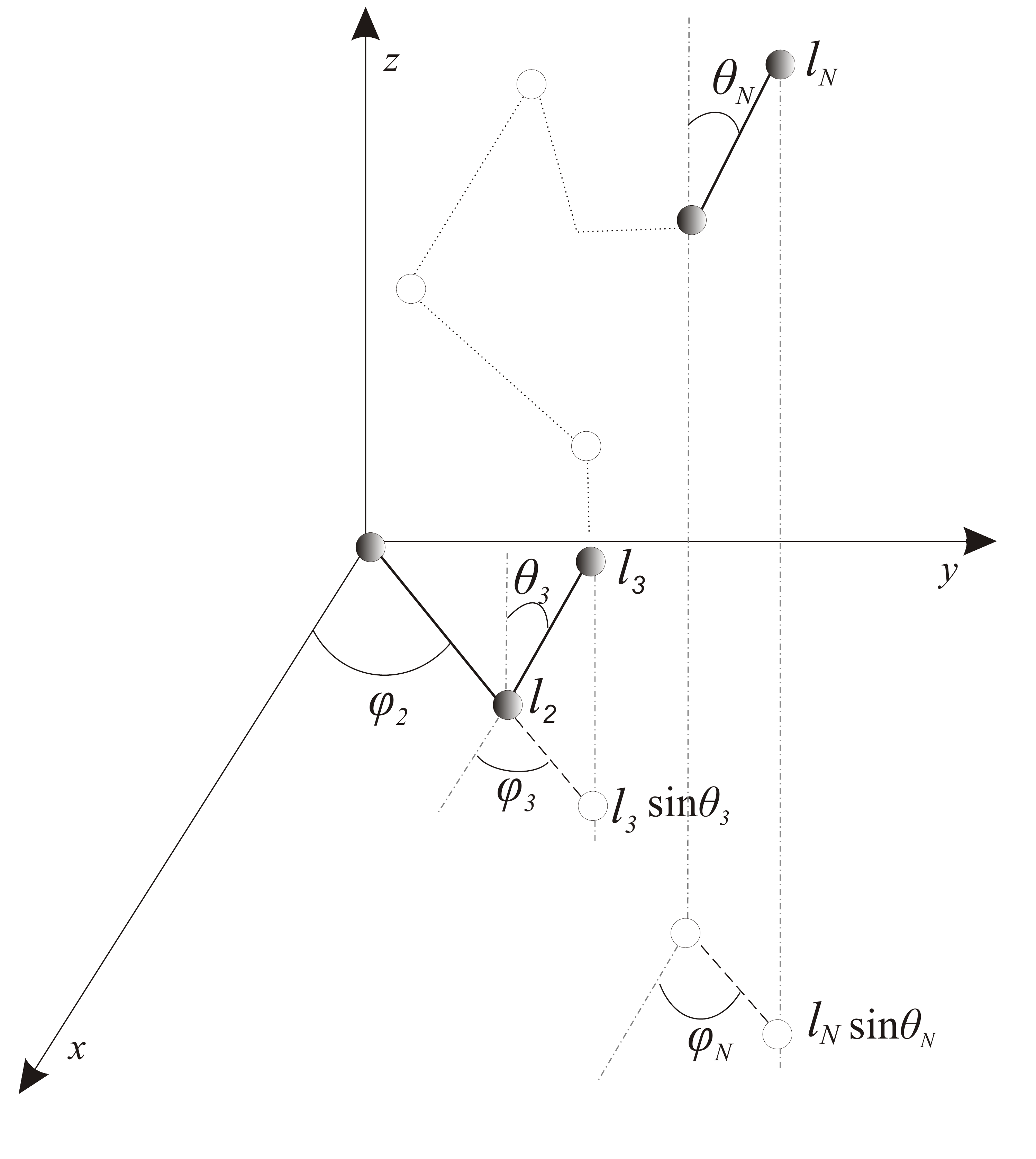}
\caption{\footnotesize Three-dimensional conformation of a chain
made of links and beads.}
 \label{system3d}
\end{figure}

The dynamics of a such a chain can be regarded as the motion of a system
of coupled
 pendulums. For the sake of simplicity
one of the ends of the chain has been fixed in the origin, see
   Fig.~\ref{system3d}.
Apart from that, no restrictions will be imposed on its motion.
This implies that different parts of the chain are allowed to penetrate
one into the other. In this case the chain is called a {\it phantom
  chain}.

The fact that
the chain is attached at the origin of the coordinates
corresponds to the condition $ P_1=(0,0,0)$ or, equivalently:
$ l_1=\dot l_1=0$.
The calculation of the kinetic energy
of the system $K_{disc}(t)$ in spherical coordinates is long but
straightforward and gives as an upshot:
\begin{eqnarray}
K_{disc}&=&\sum_{n=1}^N\sum_{k=1}^{n-1}\frac
{m_n}2l^2_{n-k+1}\dot\varphi_{n-k+1}^2\sin^2\theta_{n-k+1}
+\sum_{n=1}^{N}\sum_{k=1}^{n-1}\frac{m_n}2l^2_{n-k+1}\dot\theta^2_{n-k+1}
\nonumber\\
&+&\sum_{n=1}^{N}\sum_{k=1}^{n-1}\sum_{i=2}^{n-k}m_nl_il_{n-k+1}\left[
\dot\varphi_i\dot\varphi_{n-k+1}\sin\theta_i\sin\theta_{n-k+1}\cos\left(
\varphi_{n-k+1}-\varphi_i \right)
\right.\nonumber\\
&+&
\dot\theta_i\dot\varphi_{n-k+1}\cos\theta_i\sin\theta_{n-k+1}
\sin\left( \varphi_i-\varphi_{n-k+1}
\right)\nonumber\\
&+&\dot\varphi_i\dot\theta_{n-k+1}\sin\theta_i\cos\theta_{n-k+1}
\sin\left( \varphi_{n-k+1}-\varphi_i
\right)\nonumber\\
&+&\left.\dot\theta_i\dot\theta_{n-k+1}\left(
\cos\theta_i\cos\theta_{n-k+1}\cos\left( \varphi_{n-k+1}-\varphi_i
\right)+\sin\theta_i\sin\theta_{n-k+1} \right)\right]\label{kdisc3d}
\end{eqnarray}
To pass to the limit of a continuous chain 
we will use the rigorous procedure
 described in \cite{FEPAVI1}, where the two-dimensional case
was analyzed. In order to do that,
we assume that 
all segments have the same length and all beads have the same masses:
\beq{ l_i = a \qquad (i=2,\ldots,N) \qquad\qquad m_i = \frac{M}{L}a
\qquad (i=1,\ldots,N) }{equal2}
where $M = \sum\limits_{i=1}^N m_i$ and $L$ are
the total mass and the total length of the chain respectively. The
next step consists in performing
the limit in which the continuous system is recovered:
\beq{
a\longrightarrow 0 \qquad N\longrightarrow +\infty \qquad Na=L
}{limitcont}
One can see from \ceq{limitcont} that the product $Na$ is fixed and
gives the total length of the chain. 
Exploiting Eqs.~(\ref{kdisc3d}--\ref{limitcont})
it is possible to get the kinetic energy
 of the continuous chain. Let's see for example how the recipe
for performing the continuous limit works
 in the case of the third term in \ceq{kdisc3d}:
\begin{eqnarray}
&&\sum_{n=1}^{N}\sum_{k=1}^{n-1}\sum_{i=2}^{n-k}m_nl_il_{n-k+1}
\dot\varphi_i\dot\varphi_{n-k+1}\sin\theta_i\sin\theta_{n-k+1}\cos
(\varphi_{n-k+1}-\varphi_i)
\nonumber\\
&\longrightarrow& \frac ML
\sum_{n=1}^N a \sum_{k=1}^{n-1} a \sum_{i=2}^{n-k} a
\dot\varphi(t,s_i)\dot\varphi(t,s_n-s_k+a)
\nonumber\\
&\times&
\sin\theta(t,s_i)\sin\theta(t,s_i-s_k+a)\cos(\varphi(t,s_i-s_k+a) - \varphi(t,s_i))
\nonumber\\
&\stackrel{a\rightarrow 0}{=}&
\frac ML\int_0^Lds(L-s)\int_0^s
dv
\dot\varphi(t,v)\dot\varphi(t,s)\sin\theta(t,v)\sin\theta(t,s)
\nonumber\\
&\times&
\cos(\varphi(t,s) - \varphi(t,v))
\label{someexample}
\end{eqnarray}
In obtaining the last part of the above equation we have exploited
the formula
\beq{ \int_0^L ds \int_0^s du f(u) = \int_0^L ds (L-s) f(s)
}{formula}
 which is valid for any integrable function $f(s)$.
Applying the prescription of Eq.~\ceq{someexample} to the rest of the
terms in Eq.~\ceq{kdisc3d},
we get with the additional help of Eq.~\ceq{formula} the full
expression of the kinetic energy of the continuos chain:
\begin{eqnarray}
K(t)&=&\frac ML\int_0^Lds(L-s)\int_0^s
dv\Big[
\dot\varphi(t,v)\dot\varphi(t,s)\sin\theta(t,v)\sin\theta(t,s)
\nonumber\\
&\times& \cos\left( \varphi(t,s)-\varphi(t,v)\right)
\nonumber\\
&+&
\dot\theta(t,v)\dot\varphi(t,s)\cos\theta(t,v)\sin\theta(t,s)\sin\left
( \varphi(t,v)-\varphi(t,s)\right)\nonumber\\
&+&\dot\theta(t,s)\dot\varphi(t,v)\sin\theta(t,v)
\cos\theta(t,s)\sin\left(
\varphi(t,s)-\varphi(t,v)
\right)\nonumber\\
&+&\left.\dot\theta(t,v)\dot\theta(t,s)\Big(
\cos\theta(t,v)\cos\theta(t,s)\cos\left( \varphi(t,s)-\varphi(t,v)
\right)\right.
\nonumber\\
&+&\sin\theta(t,v)\sin\theta(t,s) 
\Big)\Big]
\label{kinene3dformone}
\end{eqnarray}
We would like to stress that the right hand side of the derived
equation contains five terms, while the
initial discrete formula of the kinetic energy
of Eq.~\ceq{kdisc3d} contained seven terms.
This is due to the fact that the contributions from Eq.~\ceq{kdisc3d}
in which $\dot\varphi^2_{n-k+1}$
and $\dot\theta^2_{n-k+1}$ are present
disappear after taking the continuous limit because they are
proportional to  $a\longrightarrow 0$.

For future convenience, we give also the expression of the kinetic
energy in cartesian coordinates:
\beq{ K_{disc}=\sum_{i=2}^N\frac {m_i}2(\dot x_i^2+ \dot
y_i^2+\dot z_i^2 ) }
{kdisc3dcartcoord}
where $x_i$, $y_i$ and $z_i$
have been defined in Eq.~\ceq{posns3d}. The sum over $i$ starts from
$2$ because one end of the chain coincides with the origin of the
axes, so that $l_1=0$. Of course, due to the condition that each
segment has a fixed length $l_i$, Eq.~\ceq{kdisc3dcartcoord} must be
completed by the following constraints: \beq{
(x_i-x_{i-1})^2+(y_i-y_{i-1})^2+(z_i-z_{i-1})^2=l_i^2\qquad\qquad
  (i=2,\ldots,N)
}{constrthreedone}

At this point we have thus two choices. Either we keep
the kinetic energy in the simple form of Eq.~\ceq{kdisc3dcartcoord}
at the price of having to deal with the constraints
\ceq{constrthreedone}, or we solve those constraints using the spherical
coordinates $l_i,\theta_i,\varphi_i$ of
Eq.~\ceq{posns3d}. In the latter case,
the kinetic energy of Eq.~\ceq{kinene3dformone} is both nonlocal and
nonlinear and thus is difficult to be treated. 
In the continuous limit, the situation does not change
substantially.

We end up this Section performing the continuous limit  of
the
kinetic energy of
Eq.~\ceq{kdisc3dcartcoord} and of the
 constraints \ceq{constrthreedone}.
Following the
prescriptions given in Eqs.~(\ref{equal2}--\ref{limitcont}),
we obtain:
\beq{ {\cal L} = \frac{M}{2L}\int_0^Lds \dot\mathbf R^2(t,s)
}{kcontthreed} and
\beq{
\mathbf R^{\prime 2} = 1
}{constrthreecontcase}
where we have introduced the vector notation:
\beq{\mathbf R = [x(t,s),y(t,s),z(t,s)]
}{bondvectors}
to describe the position on the chain.
In polymer physics $\mathbf R$ is called the {\it bond vector}.
In Eq.~\ceq{kcontthreed} and Eq.~\ceq{constrthreecontcase}
we have put
$\dot\mathbf R\equiv\frac{\partial\mathbf R}{\partial t}$ and
$\mathbf R^{\prime}\equiv\frac{\partial\mathbf R}{\partial s}$.
\\
The compatibility of the description in cartesian coordinates
with that in spherical coordinates
can
be verified
by introducing the fields
$\theta(t,s),\varphi(t,s)$ connected with
the cartesian fields $x(t,s),y(t,s),z(t,s)$ by the relations
\begin{eqnarray}
x(t,s)&=&\int_0^s
du\cos\varphi(t,u)\sin\theta(t,u)\label{contcoordtransfone}
\\
y(t,s)&=&\int_0^s du\sin\varphi(t,u)\sin\theta(t,u)
\label{contcoordtransftwo}
\\
z(t,s)&=&\int_0^s du\cos\theta(t,u) \label{contcoordtransfthree}
\end{eqnarray}
If one performs the substitutions of
Eqs.~(\ref{contcoordtransfone}--\ref{contcoordtransfthree}) in the
kinetic energy \ceq{kcontthreed} and
makes use of the formula \ceq{formula},
 one arrives exactly at the expression of the kinetic
energy \ceq{kinene3dformone}. Thus, Eq.~\ceq{kinene3dformone} and
Eq.~\ceq{kcontthreed} together with the constraint
\ceq{constrthreecontcase} are equivalent.

\section{Dynamics of a chain immersed in a heat bath}\label{s:heatbath}
In this Section the path integral formulation of an inextensible chain
in the contact with a heat reservoir
at temperature $T$
 is provided.
According to the construction  presented in
Sec.~\ref{s:class}, the conformation of the chain
is treated as the limit case of a system of $N$ beads connected by $N-1$
links of fixed length $a$. In the discrete
case the
positions of the
beads are given by a set of three-dimensional cartesian vectors
$\mathbf R_n(t)$, $(n=2,\ldots,N)$,
while the conformation of the continuous
chain at a given instant $t$ is described by the vector field $\mathbf
R(t,s)$, $s$ being the arc--length.
Furthermore, the chain is inextensible and thus
has constant
length $L=Na$.

In order to describe the thermodynamic fluctuations of the chain, we regard
it as a system of $N$
 Brownian particles of mass $m$ whose
trajectories satisfy the constraints of
Eq.~\ceq{constrthreedone}.
These constraints enforce
the condition that the total length of the links connecting the
beads should be equal to $a$.
It is possible to rewrite Eq.~\ceq{constrthreedone}
 in the more compact form:
\beq{
\frac{|\mathbf R_n(t) - \mathbf R_{n-1}(t)|^2}{a^2} = 1 \qquad\qquad (n=2,\ldots,N)
}{wiezy}
We also require that at the initial and final times
$t=0$ and $t=t_f$ the
position of $n$-th particle is
 respectively given by $\mathbf R_n(0)=\mathbf R_{0,n}$ and $\mathbf
 R_n(t_f)=\mathbf R_{f,n}$ for
 $n=2,\ldots,N$.

In other words, the primary task of this Section is
to analyze the dynamics of a system which consists
in the constrained random walk
of the beads composing the chain.
The main difficulty in performing analytical calculations
 are obviously the constraints.
Starting like in the Rouse model from an approach to the problem
based
 on the
 Langevin equation to describe the
motion of a polymer in a solution \cite{doiedwards}, the treatment
of the constraints becomes awkward.
For this reason we will use an interdisciplinary strategy, which
combines the techniques of field theory
with those used
in the statistical mechanics of polymers with topological constraints.
The starting point of the presented framework is to specify the probability
distribution function $\Psi_N$
 expressed
in a path integral form. $\Psi_N$ contains the physical information
about the system.
To be more specific, it measures the probability that the chain after
a given time $t_f$ passes from
an initial configuration $\mathbf R_{0,n}$ to a final configuration
$\mathbf R_{f,n}$.

Before we construct the probability function for the chain with rigid
constraints, let's see how the path integral
of a single free Brownian particle looks like. In order to do this we
assume that at
the time $t=0$ the particle finds itself at the initial point $\mathbf
R_0$ and starts to perform a random walk. As it is well known, the
probability $\psi(t_f;\mathbf R_f,\mathbf R_0)$ that, after the time
$t_f$ the particle arrives at a given point $\mathbf R_f$, satisfies
the diffusion equation
\beq{\frac{\partial\psi}{\partial
  t_f}=D\frac{\partial^2\psi}{\partial\mathbf R^2}}
  {basicequ}
where $D$ is the diffusion constant.
The boundary condition at $t_f=0$ is chosen in such a way that
$
\psi(0,\mathbf R_f;\mathbf R_0)=\delta(\mathbf R_f-\mathbf
R_0)
$.
The solution $\psi$ of \ceq{basicequ}  can be expressed in the form
of a path integral
\beq{
\psi(t_f,\mathbf R_f;\mathbf R_0)=A\int_{\mathbf R(0)=\mathbf
  R_0}^{\mathbf R(t_f)=\mathbf R_f}{\cal D}{\mathbf R}(t)
\exp{\left[
-\int_0^{t_f}\frac{\dot{\mathbf R}^2(t)}{4D}dt
\right]}
}{singleprobfun}
where $A$ is a normalization factor.
We note that the diffusion constant $D$ appearing in
Eq.~(\ref{singleprobfun}) satisfies the relation $D=\frac{kT\tau}{m}$,
where $k$ is the Boltzmann constant, $T$ is the temperature of
heat bath
 and $\tau$ is the relaxation time
that characterizes the rate of decay of the drift velocity of the
particle.

The above prescription can easily be generalized  to a system of $N$
noninteracting Brownian particles.
It this case
the probability that the
$n-$th particle starting from the point $\mathbf R_{0,n}$ arrives at
the point $\mathbf R_{f,n}$ is given by
\beq{
\psi_N=\prod_{n=1}^N \left[A\int_{\mathbf R_n(0)=\mathbf R_{0,n}}^{\mathbf
  R_n(t_f)=\mathbf R_{f,n}}  {\cal D}\mathbf R_n(t)\right]
\exp{\left[-\frac{1}{2kT\tau}\sum_{n=1}^N
\int_0^{t_f}\frac m2\dot{\mathbf R}_n^2(t)dt
\right]}
}{totprobfun}
In addition,
the form of the path integral on the right hand side of Eq.~\ceq{totprobfun}
displays the connection with
the partition function of a set of $N$ free particles
 in quantum mechanics where
the functional
${\cal A}_N=\sum\limits_{n=1}^N\int_{0}^{t_f}\frac m2\dot{\mathbf
  R}_n(t)dt$ represents the action of the system.
The well known duality between quantum mechanics and Brownian motions
allows to treat the factor
\beq{
\kappa=2kT\tau
}{kappa}
as the quantity which
plays the role of the Planck's constant. Indeed, one
may show that the uncertainties in the position and momentum of a
Brownian particle due to the frequent collisions with the molecules in
the solutions satisfy an analog of the Heisenberg uncertainty
relations:
$\Delta p\Delta r\sim \kappa$ \cite{rice}.

Going back to the dynamics of an inextensible chain, the only difference
with respect to a system of free particles
is that the bond vector $\mathbf R_n(t)$ satisfies the additional
constraints \ceq{wiezy}
restricting the trajectories of motion.
To implement them in the dynamics of noninteracting Brownian particles
 we add a product of functional delta functions in the path integral
 \ceq{totprobfun} which imposes the
 desired conditions \ceq{wiezy}:
\begin{eqnarray}
\Psi_{N}&=&C\left[\prod_{n=1}^N
\int_{
\mathbf R_n(0)=\mathbf R_{0,n}}
^{\mathbf R_n(t_f)=\mathbf R_{f,n}}
{\cal D}\mathbf R_n(t)\right]
e^{
-\frac M{4k_BT\tau L}
\sum_{n=1}^N a \int_{0}^{t_f}dt
\dot{\mathbf R}_n^2(t)
}\nonumber\\
&\times&\prod_{n=2}^N\delta
\left(
\frac{\left|
\mathbf R_n(t)-\mathbf R_{n-1}(t)
\right|^2}{a^2}-1
\right)
\label{partfunppint}
\end{eqnarray}
where $C$ is an irrelevant factor and the mass of a single particle
present in Eq.~\ceq{totprobfun}
has been replaced according to the equation $m=\frac ML a$.
The above  procedure to fix the constraints in a path integral has been
applied in the  statistical mechanics of entangled polymers
\cite{FKL,KV,BV}.\\
The next step consists in performing
the
continuous limit \ceq{limitcont} in
\ceq{partfunppint} \cite{kleinertpi}:
\begin{eqnarray}
\Psi=\int_{\mathbf R(0,s)=\mathbf R_0(s)}^{\mathbf
  R(t_f,s)=\mathbf R_f(s)}{\cal D}\mathbf R(t,s)e^{-\frac
  1{2kT\tau}\int_0^{t_f}dt\frac{M}{2L}\int_0^L ds\dot{\mathbf
    R}^2(t,s)}\delta (\mathbf R'^2(t,s)-1)\label{picc}
\end{eqnarray}
The result of Eq.~\ceq{picc} defines a model which is closely related
to the nonlinear sigma model
(NL$\sigma$M) used in high energy physics \cite{weinbergnlsm}, solid state physics \cite{solidnlsm}
and disordered systems \cite{disordernlsm}.
For this reason it has been called  generalized nonlinear sigma model (GNL$\sigma$M).
The most striking difference between these
 two models lays in the constraints, which in the case
of the NL$\sigma$M are of the form $\mathbf R^2=1$, while in the
GNL$\sigma$M
they have been replaced by the nonholonomic
condition \ceq{constrthreecontcase}.

To conclude this section let us note that it is possible to
show that the generating functional
of the correlation functions
of the GNL$\sigma$M coincides with the
generating functional of the correlation functions of
 the solutions of a constrained Langevin equation
\cite{FPlangevin}.

\section{Dynamics of an inextensible chain with constant bending
  angle}\label{s:3dcb}
\begin{figure}
\centering
\includegraphics[scale=0.4]{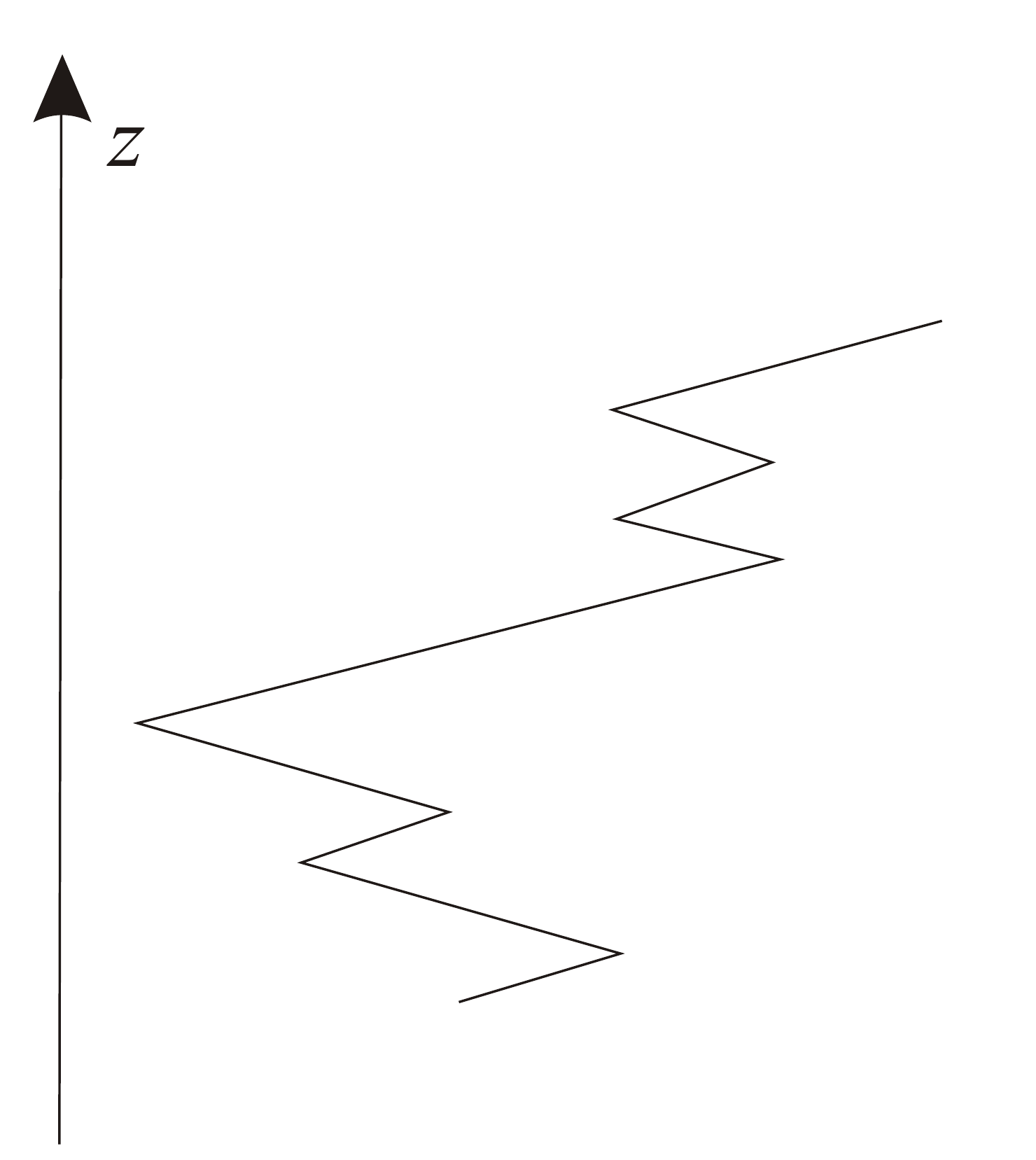}
\caption{Example of motion of a chain whose segments are constrained
  to form a fixed angle $\alpha$ with the $z-$axis. In the figure
  $\alpha=30^\circ$.} \label{motconang}
\end{figure}

The approach presented in Sec.~\ref{s:class} in order to treat the dynamics of
random chains has some interesting variants which we would like to
discuss in this Section. To this purpose, we choose the formulation
in which the positions of the ends of the segments composing the
chain are given in cartesian coordinates.
As we have already seen,
in this way the expression of the kinetic energy $K_{disc}$ is given
by \ceq{kdisc3dcartcoord} and
must be completed by the constraints \ceq{constrthreedone}.
From now on we assume as before that all segments
have the same fixed length $l_n=a$, but  we require additionally
that:
\beq{ (z_n-z_{n-1})^2=b^2\le a^2 }{addrequir}
This implies
that the projection of each segment onto the $z-$axis has length
$\pm b$, so that the segments are bound to form with the $z-$axis
the fixed angles $\alpha_1=\alpha$ or $\alpha_2= (\pi-\alpha)$
defined by the relations:
\beq{\cos\alpha_1=+\frac ba\qquad\qquad
\cos\alpha_2=-\frac
  ba}{anglealphadef}
Clearly, in both cases the constraints \ceq{constrthreedone} and
\ceq{addrequir} may be rewritten as follows:
\beq{
\frac{(x_n-x_{n-1})^2}{b^2}+\frac{(y_n-y_{n-1})^2}{b^2}=\frac
1{\cos^2\alpha}-1\qquad\qquad (n=2,\ldots,N) }{constralpha}
where $\alpha$ may be either $\alpha_1$ or $\alpha_2$.
In the following we will suppose
 that
 only the angle
$\alpha=\alpha_1$ is allowed, so that the chain cannot make turns in the
$z$ direction. An example of a conformation of a chain satisfying
these assumptions is given in Fig.~\ref{motconang}.

The constraints \ceq{constralpha} are eliminated using the spherical
coordinates of Eq.~\ceq{posns3d} after setting
the angles
$\theta_n$ formed by the
segments with the $z-$axis equal to $\alpha$:
\begin{eqnarray}
x_n(t)&=&\sum_{i=1}^nl_i\cos\varphi_i(t)\sin\alpha\label{varonedf}\\
y_n(t)&=&\sum_{i=1}^nl_i\sin\varphi_i(t)\sin\alpha\label{vartwodf}\\
z_n(t)&=&\sum_{i=1}^nl_i\cos\alpha=na\cos\alpha\label{coordsab}
\end{eqnarray}
As we see from the above equation, each segment is left  only with
the freedom of rotations around the $z-$direction, corresponding to
the angles $\varphi_i(t)$. Moreover, the total length of the chain
is always $L=Na$, but now also the total height $h$ of the
trajectory along the $z-$axis is fixed: \beq{h=Nb}{totalheightz} At
this point, we pass to the continuous limit, this time taking as
parameter describing the trajectory of the chain the variable $z$
instead of the arc-length $s$. Due to the last of
Eqs.~\ceq{coordsab}, the $z-$components of the velocities are always
zero: \beq{\dot z_n(t)=0}{constzdot} As a consequence, we are left
with something similar to a two-dimensional problem.
The  difference
from a real two-dimensional problem, which could be obtained by  putting
$\theta_j=\pi/2$
($j=1,\ldots,N$) in Eq.~\ceq{posns3d},
is that the equations describing the position of a bead in
two dimensions, namely
$x_i(t)=\sum_{j=1}^Nl_j\cos\varphi_j(t)$ and
$y_i(t)=\sum_{j=1}^Nl_j\sin\varphi_j(t)$,
have been replaced by Eqs.~\ceq{varonedf} and \ceq{vartwodf}. Moreover,
the
constraints have a slightly different form. Following the same
procedure presented in Sec.~\ref{s:class}, we find after a few
calculations the expression of the kinetic energy in the continuous limit:
\begin{eqnarray}
K_{\alpha}&=&\tan^2\alpha\frac M{2h}\int_0^hdz\int_0^zdz_1\int_0^{z_1}dz_2
\nonumber\\
&\times& \dot\varphi(t,z-z_1)\dot\varphi(t,z_2) \cos(
\varphi(t,z-z_1) -\varphi(t,z_2))\label{kcont3dalphaone}
\end{eqnarray}
and of the constraint \ceq{constralpha}: \beq{ (\partial_z
x)^2+(\partial_z y)^2=\tan^2\alpha }{constralphacont}
It is also not
difficult to show that the probability distribution
$\Psi_{\alpha}
$ is given in cartesian coordinates by: \beq{ \Psi_{\alpha}
= \int{\cal D}x(t,z){\cal D}y(t,z) \exp\left\{-\frac{{\cal
A}_{\alpha}}{\kappa} \right\}\delta( (\partial_z x)^2+(\partial_z
y)^2-\tan^2\alpha ) }{3ddirfin} where \beq{ {\cal
A}_{\alpha}=\frac{M}{2h}\int_{0}^{t_f}dt\int_0^hdz\left[ \dot
x^2(t,z)+\dot y^2(t,z) \right] }{djfjskdflds}

\begin{figure}
\centering
\includegraphics[scale=0.37]{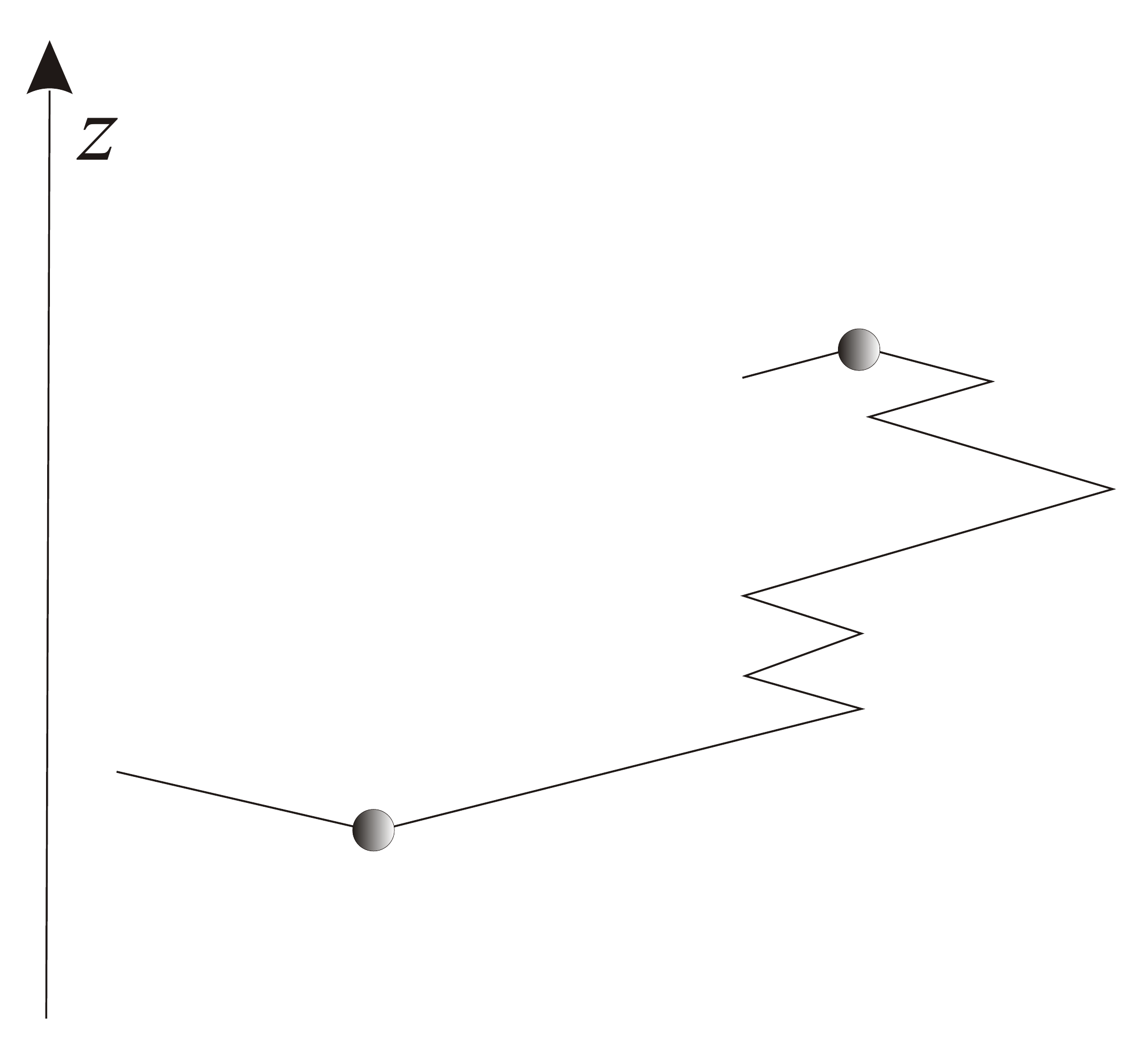}
\caption { Example of motion of a chain whose segments are constrained
  to form fixed angles $\alpha$ or
$\pi-\alpha$
 with the $z-$axis. In the figure
  $\alpha=30^\circ$.  Turning points are emphasized by means of shaded beads.}
  \label{turnedchain}
\end{figure}

At this point we discuss briefly the case in which both angles
$\pi-\alpha$ and $\alpha$ are allowed. In this situation, the
trajectory of the chain may have turns. An example of motion of this
kind is given in Fig.~\ref{turnedchain}.

The constraints \ceq{constrthreedone} and \ceq{addrequir} remain
unchanged, but the coordinate $z$ cannot be chosen as a valid
 parameter of the trajectory of the chain and
one has to come back to the arc-length $s$. The most serious problem
is the fact that the variables $z_n(t)$ are not
continuous functions of the time, since the length $z_n(t)-z_{n-1}(t)$
is allowed to
jump discretely between the two discrete values $+b$ and $-b$,
corresponding to the angles $\alpha$ and $\pi-\alpha$ respectively.
It is therefore difficult to define the components $\dot z_n$ of the
velocities of the ends of the segments and thus their contribution
to the  kinetic energy. Let us note that this problem affects only
the $z$ degrees of freedom. The degrees of freedom $x_n(t)$ and
$y_n(t)$ of the chain remain continuous functions of $t$ despite the
jumps of the $z_n$'s. This fact can be easily verified looking at
the definition of $x_n(t)$ and $y_n(t)$ in Eqs.~\ceq{varonedf} and
\ceq{vartwodf}. Since $\sin(\pi-\alpha)=\sin\alpha$, both the
$x_n(t)$'s and $y_n(t)$'s are not affected by the jumps of the angle
$\alpha\longleftrightarrow \pi-\alpha$. As a consequence, the
problems with the $z$
variable can easily be solved
if the chain has no interactions in which the $z$ variable is
involved. In this case, in fact,
the degrees of freedom connected to the motion along the
$z-$direction
are decoupled from the other degrees of freedom and may be neglected.

As a consequence, we assume that the interactions are $z-$independent,
so that the difficulties related to the motion along the
$z-$direction disappear and once again the problem reduces to that
that of a two-dimensional chain. Since the
constraints are always those of
Eqs.~(\ref{constrthreedone}) and (\ref{addrequir}), one may proceed as in
the case of fixed angle $\alpha$. As a result, one finds that the
final probability distribution is of the form:
\begin{eqnarray}
\Psi_{\alpha,\pi-\alpha}
&=&\nonumber\\
&&\!\!\!\!\!\!\!\!\!\!\!\!\!\!\!\!\!\!\!\!\!\!\!\!\!\!\!\!
\!\!\!\!\!\!\!\!\!\!\!\!\!\!\!\!\!\!\!\!\!\!\!\!\!\!\!\!\!\!\!\!
{\cal C}\int{\cal D}x(t,s){\cal D}y(t,s) \exp\left\{-\frac{{\cal
    A}_{\alpha,\pi-\alpha}}{\kappa}
\right\}\delta( (\partial_s x)^2+(\partial_s y)^2-\tan^2\alpha )
\label{3ddirfiwn}
\end{eqnarray}
where \beq{ {\cal
  A}_{\alpha,\pi-\alpha}=\frac{M}{2L}\int_{0}^{t_f}
dt\int_0^Lds\left[ \dot x^2(t,s)+\dot y^2(t,s) \right] }{djfjskdfldws} and
${\cal C}$ is a constant containing the result of the integration
over the decoupled $z$ degrees of freedom. With respect to the
previous case, let
us note that in
Eqs.~\ceq{3ddirfiwn} and \ceq{djfjskdfldws}
 $z$ has been replaced by the arc-length $s$ as the parameter of
the trajectory of the chain. Correspondingly, the total chain length
$L$ appears in the action instead of the height $h$.

\section{Conclusions}\label{conclusions}
In this work the dynamics of an inextensible
freely jointed chain consisting of
links and beads in three dimensions
has been discussed both from the
classical and statistical point of view.
In Sec.~\ref{s:class} we have mainly
concentrated ourselves on the computation of the kinetic energy
in cartesian and spherical coordinates for the discrete and continuous chain.
In Sec.~\ref{s:heatbath} we have derived the probability function of the chain $\Psi$   using a path integral framework and the fact that
the fluctuations of the chain can be regarded 
as those of a system of Brownian  particles with an additional constraint condition
 imposed
 on their trajectories.
The probability function $\Psi$  of
this system  is
equivalent to the partition function of a generalized nonlinear $\sigma$
model.
The analogy of the GNL$\sigma$M with the NL$\sigma$M suggests the possibility of applying techniques and results coming from the NL$\sigma$M
to the GNL$\sigma$M. For example, it is known that the NL$\sigma$M is renormalizable in two dimensions and also that it has interesting
features because it  is analytically free and has a dynamically generated mass gap \cite{andersen}.
The similarity with NL$\sigma$M seems also to suggest that there is no symmetry breaking in the underlying $O(d)$ symmetry of the GNL$\sigma$M,
where $d$ denotes the dimension of the vector field $\mathbf R(t,s)$. One should however be careful when extending the results of the NL$\sigma$M
to the GNL$\sigma$M. For example if $d=2$, one may use polar field coordinates to express the vector field $\mathbf R(t,s)$. If one does that
the NL$\sigma$M becomes a free field theory in the angle variable \cite{zinn}.
 This is not true in the case of the GNL$\sigma$M which in polar coordinates exhibits a
nonlinear and complicated dependence on the angle degree of freedom. Moreover, it is not straightforward to apply techniques like the
effective potential method which is useful to investigate possible phase transitions  in the NL$\sigma$M. The reason is that in this method it is
performed an expansion around field configurations minimalizing the action which are constant. Configurations of this kind correspond in the
GNL$\sigma$M to the situation in which the chain has collapsed to a point and thus are nonphysical.
Finally in Sec.~\ref{s:3dcb}  a three-dimensional chain admitting only fixed angles
with respect to the $z-$axis has been discussed. It has been shown that it is possible
to reduce the problem to two dimensions, in a way which is similar to the reduction of the
statistical mechanics of a directed polymer to the random walk of a
two-dimensional particle \cite{kamien}.
Our approach is valid only if the chain has no turning points. If
there are turning points
 the kinetic energy is not well defined, because the variable
$z(t,s)$ is no longer a continuous function and thus its time
derivative becomes a distribution. One way for adding to our
treatment turning points as those of Fig.~\ref{turnedchain} is to
replace the variable $z$ with a stochastic variable which is allowed
to take only discrete values. Another way is to look at turning
points as points in which the chain bounces against an invisible
obstacle. A field theory describing a one-dimensional chain with
such kind of constraints has been already derived in
Refs.~\cite{arodz}.

\section{Acknowledgements}
This work has been financed by the Polish Ministry of Science and
Higher Education, scientific project N202 156 31/2933.
The authors wish to thank the anonymous referee for useful comments.


\begin{thebibliography}{99}

\bibitem{EdwGoo1} S. F. Edwards and A. G. Goodyear, {\it J. Phys. A:
  Gen. Phys.} {\bf 5} (1972), 965.
\bibitem{EdwGoo2}  S. F. Edwards and A. G. Goodyear, {\it J. Phys. A:
  Gen. Phys.} {\bf 5} (1972), 1188.
\bibitem{EdwGoo3}  S. F. Edwards and A. G. Goodyear, {\it J. Phys. A:
  Gen. Phys.} {\bf 6} (1973), L31.

\bibitem{ottinger} H. C. \"Ottinger, {\it Phys. Rev.}
 {\bf E 50} (1994), 2696.
\bibitem{muthukumar} D. Petera and M. Muthukumar,
{ \it J. Chem. Phys.} {\bf 111} (1999), 7614.
\bibitem{adland} H. M. {\AA}dland and A. Mikkelsen,
 {\it J. Chem. Phys.} {\bf 120} (2004), 9848.
\bibitem{FEPAVI1} F. Ferrari, J. Paturej and T. A. Vilgis, {\it  Phys. Rev.} {\bf E 77} (2008), 021802.
\bibitem{FPlangevin} F. Ferrari and J. Paturej, {\it J. Phys. A} {\bf 42}, 145002.
\bibitem{FEPAVI2}  F. Ferrari, J. Paturej and T. A. Vilgis,
{\it Applications of a generalization of the nonlinear sigma model
  with $O(d)$ group of symmetry to
the dynamics of a constrained chain}, {\bf arXiv:0807.4045} (submitted to {\it Nucl. Phys.} {\bf B})
\bibitem{FEPAVI3} F. Ferrari,  J. Paturej, T. A. Vilgis and T. Wydro,
{\it The probability distribution of the average relative distance
  between two points in a dynamical chain}, {\bf arXiv:0809.2261} (submitted to {\it J. Chem. Phys.})

\bibitem{doiedwards} M. Doi and S.F. Edwards, {\sc The Theory of Polymer
Dynamics}
  (Clarendon Press, Oxford, 1986).



\bibitem{rice} S. A. Rice and H. L. Frisch, {\it
  Ann. Rev. Phys. Chem.} {\bf 11} (1960), 187.

\bibitem{FKL} F. Ferrari, H. Kleinert and I. Lazzizzera, {\it Int. Jour. Mod. Phys.}
 {\bf B 14} (1998), 3881.
\bibitem{KV} A. L. Kholodenko and T. A. Vilgis, {\it{Phys. Rep.}} {\bf{298}} (1998), 251.
\bibitem{BV} M. G. Brereton and T. A. Vilgis,  {\it J. Phys. A: Math. Gen.}  {\bf 28} (1995), 1149.
\bibitem{kleinertpi} H. Kleinert, {\sc Gauge Fields in Condensed Matter} (World Scientific, Singapore, 1990), Vol. 1.

\bibitem{weinbergnlsm} S. Weinberg, {\it Phys. Rev. Lett.} {\bf 18} (1967), 188; {\it Phys. Rev.} {\bf 166} (1968), 1568.
\bibitem{solidnlsm} P. B. Wiegmann, {\it J. Phys. C: Solid State Phys.} {\bf 11} (1978), 1583;
F. D. Haldane, {\it Phys. Rev. Lett.} {\bf 61} (1988), 1029.
\bibitem{disordernlsm} V. R. Kogan, K. B. Efetov,
{\it Phys. Rev.} {\bf B 67} (2003), 245312;
K. Takahashi, {\it Phys. Rev.} {\bf E 70} (2004), 066147.
\bibitem{andersen}J. O. Andersen, D. Boer and H. J. Warringa, {\it Phys. Rev.} {\bf D 69} (2004), 076006.
\bibitem{zinn} J. Zinn-Justin, {\sc Quantum Field Theory and Critical
Phenomena},
  (Clarendon Press, Oxford, 2002).
\bibitem{kamien} R. D. Kamien, P. Le Doussal and D. R. Nelson, {\it Phys. Rev.} {\bf A 45} (1992), 8727.
\bibitem{arodz} H. Arod\'z, P. Klimas and T.
  Tyranowski, {\it Phys. Rev.} {\bf E 73} (2006); {\it Acta Phys. Pol.} {\bf B 38} (2007), 2537;
{\it Acta Phys. Pol.} {\bf B 36} (2005), 3861; H. Arod\'z, {\it
    Acta Phys. Pol.} {\bf B 33} (2002), 1241; H. Arod\'z, {\it Acta Phys.
    Pol.} {\bf B 35} (2004), 625.
 \end{thebibliography}
\end{document}